\documentclass[prl,twocolumn,superscriptaddress,floats]{revtex4}

\usepackage[dvips]{graphicx}
\usepackage{amsmath}
\usepackage{booktabs}
\usepackage{dcolumn}

\newcolumntype{d}{D{.}{.}{-1}}

\newcommand{\tzg}{t$_{2g}$}

\newcommand{\sro}{Sr$_2$RuO$_4$}
\newcommand{\srdzs}{Sr$_3$Ru$_2$O$_7$}

\newcommand{\cro}{Ca$_2$RuO$_4$}
\newcommand{\csrx}{Ca$_{2-x}$Sr$_{x}$RuO$_4$}
\newcommand{\csrea}{Ca$_{1.8}$Sr$_{0.2}$RuO$_4$}
\newcommand{\csref}{Ca$_{1.5}$Sr$_{0.5}$RuO$_4$}
\newcommand{\csreda}{Ca$_{1.38}$Sr$_{0.62}$RuO$_4$}
\newcommand{\vq}{${{\bf q}}$}

\newcommand{\vQ}{${{\bf Q}}$}
\newcommand{\gbd}{$\gamma$ band}
\newcommand{\abbd}{$\alpha , \beta$ band}

\begin{document}

\textheight 24.6 true cm

\title{Field-induced paramagnons at the metamagnetic transition in Ca$_{1.8}$Sr$_{0.2}$RuO$_4$ }

\author{P. Steffens }
\affiliation{II. Physikalisches Institut, Universit\"{a}t zu K\"{o}ln, Z\"{u}lpicher Str. 77, D-50937 K\"{o}ln, Germany}

\author{Y. Sidis}
\affiliation{Laboratoire L\'eon Brillouin, C.E.A./C.N.R.S., F-91191 Gif-sur-Yvette CEDEX, France}

\author{P. Link}
\thanks{Also at : Spektrometer PANDA, Institut f\"ur Festk\"orperphysik, TU Dresden}
\affiliation{ Forschungsneutronenquelle Heinz Maier-Leibnitz (FRM-II), TU M\"unchen, Lichtenbergstr. 1, 85747 Garching, Germany}

\author{K. Schmalzl}
\affiliation{Institut Laue Langevin, 6 Rue Jules Horowitz BP 156, F-38042 Grenoble CEDEX 9, France}

\author{S. Nakatsuji}
\affiliation{Institute for Solid State Physics, University of Tokyo, Kashiwa, Chiba 277-8581, Japan.}

\author{Y. Maeno}
\affiliation{Department of Physics, Kyoto University, Kyoto 606-8502, Japan}

\author{M. Braden}
\affiliation{II. Physikalisches Institut, Universit\"{a}t zu K\"{o}ln, Z\"{u}lpicher Str. 77, D-50937 K\"{o}ln, Germany}

\date{\today}

\pacs{PACS numbers:78.70.Nx, 74.70.Pq, 75.40.Gb}

\begin{abstract}

The magnetic excitations in Ca$_{1.8}$Sr$_{0.2}$RuO$_4$ were studied across the metamagnetic transition and as a function of temperature
using inelastic neutron scattering. At low temperature and low magnetic field the magnetic response is dominated by a complex
superposition of incommensurate antiferromagnetic fluctuations. Upon increasing the magnetic field across the metamagnetic transition,
paramagnon and finally well-defined magnon scattering is induced, partially suppressing the incommensurate signals. The high-field phase in
Ca$_{1.8}$Sr$_{0.2}$RuO$_4$ has, therefore, to be considered as an intrinsically ferromagnetic state stabilized by the magnetic field.

\end{abstract}

\maketitle

\begin{figure}[t]
\begin{center}
\includegraphics*[width=0.85\columnwidth]{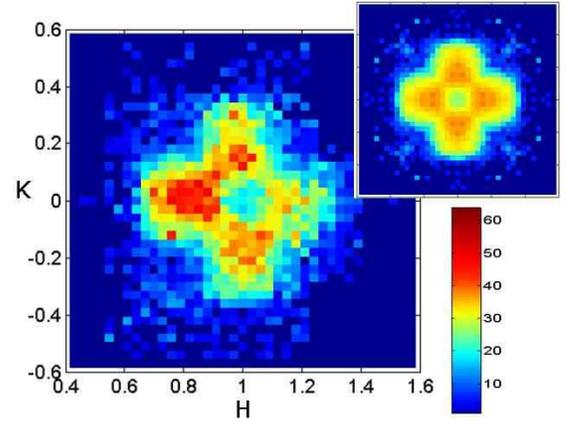}
\end{center}
\caption{(color online) Mapping of the magnetic intensity in the \textit{a*}/\textit{b*}-plane of reciprocal space around
\textbf{Q}=(1,0,0). Data are taken at T=2~K and at an energy transfer of 2.5~meV. A smooth background is subtracted. The inset shows the
same data, but fully symmetrized and corrected for the magnetic form factor $F(\textbf{Q})$. } \label{fig1}
\end{figure}

\begin{figure}[t]
\begin{center}
\includegraphics*[width=1\columnwidth]{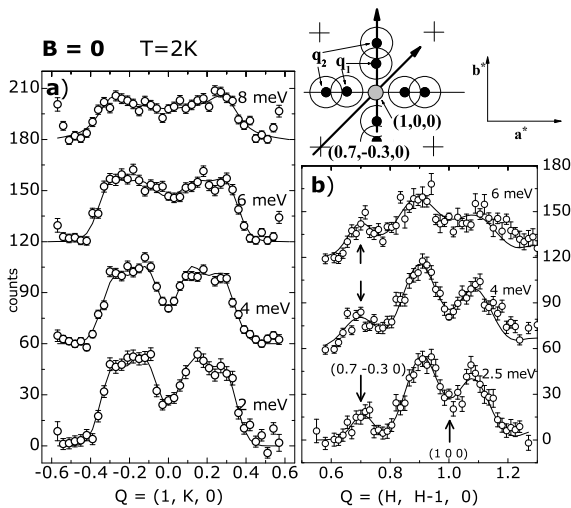}
\end{center}
\caption{Constant energy scans: (a) along $b^\ast$, (b) along the diagonal of the Brillouin zone. The sketch on top illustrates the
positions of the incommensurate magnetic signals as assumed in the model described in the text, and the bold arrows are the trajectories
of the scans in reciprocal space. The crosses are the weaker $\alpha$/$\beta$ nesting signal at (1,0,0) + ($\pm$0.3,$\pm$0.3,0). The lines
are fits with Gaussians centered at the positions shown in the sketch.} \label{fig2}
\end{figure}

Metamagnetic transitions have recently attracted considerable interest as, despite their typically first order character, a quantum phase
transition can be realized when the critical end point is driven to zero temperature. By varying the direction of the applied field, this
suppression of the critical end point can be achieved in the double layer ruthenate \srdzs\ inducing fascinating quantum-critical
phenomena \cite{grigera}.

\csrea\ belongs to the single layered ruthenates but exhibits a metamagnetic transition very similar to that in \srdzs.  In the series
\csrx\ \cite{nakatsuji00}, severe structural distortions induce a wide variety of  physical properties between the Mott insulator \cro\
and the unconventional superconductor \sro. \csrea\ is a metal with a very high electronic specific heat coefficient well in the range of
typical heavy fermion compounds, indicating strong magnetic fluctuations \cite{nakatsuji03}. At magnetic fields of 2-8T (depending on the
field orientation) it undergoes the metamagnetic transition into a state with high magnetic polarization of about 0.7$\mu _B$ per Ru
\cite{nakatsuji03} accompanied by a shift in the occupation of the Ru 4d \tzg -states \cite{kriener,balicas}. However, it is still an open
issue whether the high-field phase is intrinsically ferromagnetic or just polarized, in \csrea \ as well as in \srdzs . Due to the
intrinsic disorder caused by the chemical doping, the metamagnetic transition in \csrea\ is smeared out. In consequence, any
quantum-critical scaling might be strongly modified \cite{joerg}. However, the simpler crystal structure with only one RuO$_2$ layer and
the resulting simpler (essentially 2-dimensional) electronic band structure render \csrea\ more favorable for an analysis of the
underlying magnetic mechanism. Previous studies on related ruthenates, \csrx\ with x=0.62 \cite{friedtprl} and \srdzs\
\cite{capogna,stone}, have revealed the complexity of the magnetic response at zero field but did not address the metamagnetic transition.

The cross section for magnetic inelastic neutron scattering (INS) is given by the imaginary part of the susceptibility
$\chi''(\textbf{Q},\omega)$ \cite{lovesey}: $\frac{d^2\sigma}{d\Omega d\omega} \propto
\frac{F^2(\textbf{Q})}{1-\exp(-\frac{\hbar\omega}{k_BT})}\cdot \chi''(\textbf{Q},\omega)$. $F(\textbf{Q})$ is the magnetic form factor. We
used two coaligned single crystals of \csrea\ of about 3~mm diameter and 15~mm length each, grown at Kyoto University. The measurements
were performed on different neutron triple-axis spectrometers: 2T and 4F at the LLB, Saclay, IN22 at the ILL, Grenoble, and PANDA at
FRM-2, Garching. On the two latter spectrometers, we applied magnetic fields up to 10T using vertical cryomagnets, i.e.\ perpendicular to
the scattering plane. Two different sample orientations were used: one with the \textit{a}- and \textit{b}-axis, and one with the
\textit{a}- and \textit{c}-axis in the scattering plane. Throughout this article, we use the pseudo-tetragonal notation $a=b=3.76\
\text{\AA}$ and $c=12.55\ \text{\AA}$ in accordance with the majority of the literature, neglecting the structural distortions causing an
orthorhombic unit cell, $\sqrt{2} a \times \sqrt{2} a \times 2 c$ \cite{friedtprb}. The sample was twinned, with approximately equal
amounts of both twins.

\begin{figure}[t]
\begin{center}
\includegraphics*[width=0.9\columnwidth]{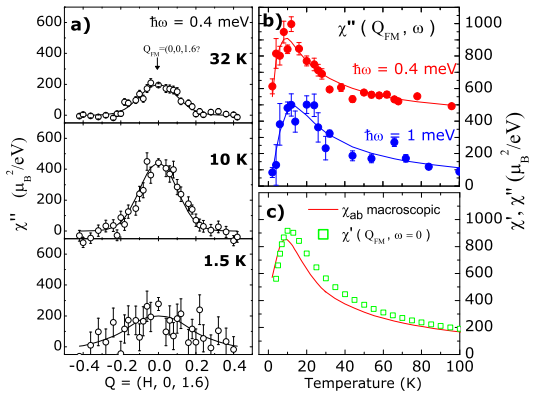}
\end{center}
\caption{(color online) Temperature dependence of the paramagnon scattering. (a):  Constant energy scans at three different temperatures.
In (b) we plot the intensity of the (2D) zone center Q$_\text{FM}$=(0,0,1.6) as function of temperature for 0.4 and 1~meV energy transfer
(lines are guides to the eye, 0.4~meV-data shifted by 400$\frac{\mu_B^2}{\text{eV}}$). From the smoothed curves in (b) we estimate the real
part of the susceptibility at $\omega=0$ (see text), which is plot in (c) together with the macroscopic susceptibility (in-plane average)
>from reference \cite{nakatsuji00}. } \label{fm-contr}
\end{figure}

\begin{figure}[t]
\begin{center}
\includegraphics*[width=0.95\columnwidth]{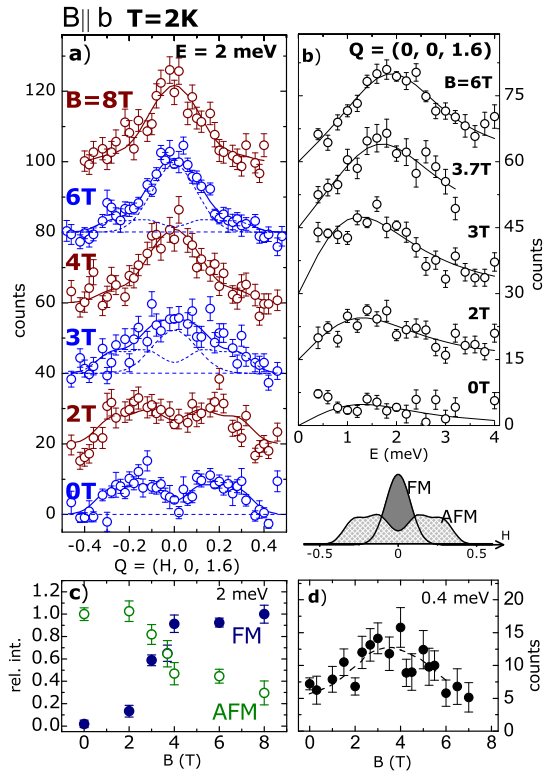}
\end{center}
\caption{(color online) Magnetic scattering for different magnetic fields applied along the b-axis:  (a) Constant energy scans along [100].
The lines assume a model consisting of an antiferromagnetic and a ferromagnetic (centered at Q$_{FM}$=(0,0,1.6)) contribution, as sketched
in the figure. At 3 and 6~T, the contributions are shown separately. The variation of the scale factors of these two contributions
relative to their values at 0 resp. 8~T is shown in (c) as function of field. In (b) we show energy scans on Q$_{FM}$, and (d) summarizes
the field dependence of the low energy part on Q$_{FM}$. (Lines in (b),(d) are guide to the eye.)} \label{fig3}
\end{figure}

\begin{figure}[t]
\begin{center}
\includegraphics*[width=0.95\columnwidth]{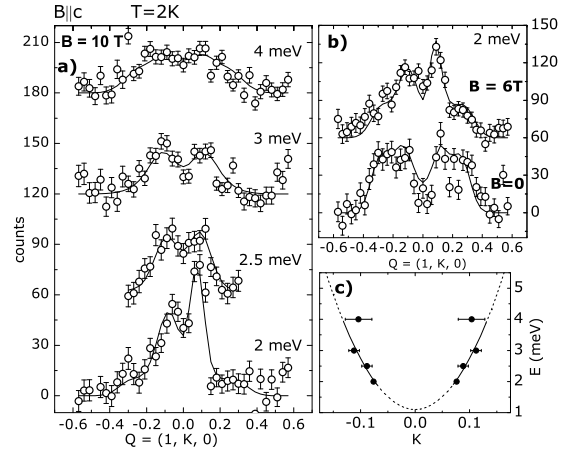}
\end{center}
\caption{(a): Constant energy scans above the metamagnetic transition (B=10~T, B$\parallel$c)  in the same configuration as the scans in
Figure \ref{fig2}. The lines are fits consisting of two contributions: the antiferromagnetic one was taken to be the same as in Fig.\
\ref{fig2}a times a variable scale factor $\leq 1$; the magnon was added as two Gaussians at symmetric positions, but variable width to
account for focusing effects of the spectrometer. In (b) the 2~meV scans at 0 and 6~T are plot for comparison. In (c), the fitted peak
positions are plot together with a parabolic dispersion.} \label{fig5}
\end{figure}

Let us first consider the spin dynamics in \csrea\ at zero magnetic field and low temperature, T=2\ K. Figure \ref{fig1} shows the neutron
intensity at constant energy transfer of $\hbar\omega=2.5$ meV mapped out in reciprocal space around \textbf{Q}=(1,0,0). For the
description of the magnetism we restrict ourselves on the two dimensions formed by the RuO$_{2}$-plane (\textit{ab} plane), because the
correlation of adjacent planes, i.e.\ along the \textit{c} axis, is negligible \cite{sidis,friedtprl}. Therefore, \textbf{Q}=(1,0,0) can be
regarded as a ferromagnetic (FM) zone center, and the area shown in Fig.\ \ref{fig1} covers already a full Brillouin zone. The magnetic
scattering is broadly distributed around the FM zone center  resembling that observed in \csreda\ \cite{friedtprl}, where the metamagnetic
transition is strongly suppressed. At the temperature of 2\ K in \csrea , we always find a minimum at the FM center (1,0,0) in cuts along
the [100] or [110] directions (Fig.~\ref{fig1} and Fig.~\ref{fig2}a,b). Furthermore, the data cannot be described by a single contribution
centered on the \textit{a*}/\textit{b*} axes; the scans along [010] show steep edges at \vQ=(1,\ $\pm0.35$,\ 0) and a relatively broad and
flat plateau between (1,\ $\pm0.1$,\ 0) and (1,\ $\pm0.3$,\ 0). For a phenomenological description, we fit the magnetic scattering  by
\emph{two} Gaussian contributions on each side of (1,0,0) with approximately equal intensity and width. These contributions are centered
at \vq$_1$= (1,\ 0.12$\pm$0.01,\ 0) and \vq$_2$=(1,\ 0.27$\pm0.01$,\ 0) and the equivalent positions in pseudo-tetragonal symmetry. We may
exclude that the inner signal stems from an isotropic paramagnon signal, as there is no evidence for a ring of scattering in Fig. 1 and as
the intensity does not show the expected  increase  \cite{lovesey} when approaching (1,0,0) and low energies. There is no visible shift of
the \vq$_1$,\vq$_2$ signals as function of energy, but rather a broadening which finally suppresses the minimum at the center. In
agreement with the study on \csreda \ \cite{friedtprl} the \vq$_1$,\vq$_2$ fluctuations exhibit a characteristic energy of 2.7$\pm$0.2\
meV. From the analysis of the geometrical factor at different equivalent \textbf{Q}-points, we deduce that for these fluctuations
$\chi''_{c}$ is significantly smaller than $\chi''_{ab}$ which is opposite to the finding in \sro \ \cite{pol-srruo}.

In the layered ruthenates, the Fermi surface consists of several sheets related with the three \tzg -states of the Ru 4d-shell
\cite{band}: two sheets arise from nearly one-dimensional bands of $d_{xz/yz}$ character, named $\alpha$ and $\beta$, and the
$\gamma$-sheet originates from the two-dimensional 4d$_{xy}$ band. We assume that the \vq$_1$,\vq$_2$ signals arise from the \gbd , since
this band has been shown to carry the magnetization in \csref\ \cite{gukasov}, although its topology (electron- or hole-like \cite{wang})
in the structurally distorted compound is not fully established yet.

The magnetic excitations described above strongly resemble  those observed in  \srdzs\ with two incommensurate contributions on the
\textit{a*}/\textit{b*} axes at almost the same positions (x=0.09 and 0.25) \cite{capogna}. This remarkable agreement suggests that
magnetic properties and, in particular, the metamagnetic mechanism should be very similar in these ruthenates.

The diagonal scans in Fig.\ \ref{fig2}b show a weaker signal at $\bf{Q}_{(\alpha\beta)}\approx$(0.7,-0.3,0), at the position where the
dominant magnetic fluctuations occur in \sro \ due to nesting in the $\alpha$ and $\beta$ Fermi-surface sheets \cite{sidis,band}. The
structural distortions in \csrea\ cause a folding of the electronic bands with respect to (0.5,0.5,0) implying a complex Fermi surface as
well as a complex Lindhard function. However, the strong nesting tendency observed in \sro \ should be rather robust and can be taken as a
rough estimate of the band filling. In \csrea\ the filling of the \abbd s seems thus to be similar to that in \sro \ clearly contradicting
the proposal of an orbital-selective Mott transition \cite{anisimov} requiring a significant redistribution of orbital occupation.

Upon heating to intermediate temperatures of the order of 10~K, the magnetic response changes significantly developing strong paramagnon
scattering of truly FM character on top of the already complex low-temperature AFM response described above. Upon cooling, the paramagnon
contribution appears below $\sim$50~K, passes a maximum and becomes suppressed at the lowest temperatures, see Fig.\ \ref{fm-contr}. The
paramagnon scattering exhibits a significantly smaller characteristic energy of the order of  0.2-0.6\ meV which strongly depends on
temperature. Via calibration by an acoustic phonon, we may determine $\chi''(Q,\omega)$ in absolute units, see Fig.\ \ref{fm-contr}, which
then allows us to calculate $\chi'(Q,0)$ by Kramers-Kronig analysis. For a single relaxor
$\chi''(Q,\omega)=\frac{\chi'(Q,0)\cdot\Gamma\cdot\omega}{\Gamma^2+\omega^2}$, $\chi'(Q,0)$ amounts to twice the maximum of the imaginary
part (at $\omega=\Gamma$). Due to the strong in-plane anisotropy, $\chi''(Q_\text{FM},\omega)$ consists of two contributions, and the
average $\chi''(Q_\text{FM},0.4\text{meV})+\chi''(Q_\text{FM},1\text{meV})$ is a reasonable estimate for $\chi'(Q_\text{FM},0)$. Figure\
\ref{fm-contr}c compares this estimate with the susceptibility measured by a macroscopic method yielding good agreement. The incipient FM
instability in \csrea, which due to the orbital rearrangement \cite{kriener,balicas,joerg} is suppressed at low temperature, is thus
carried by the paramagnon fluctuations.

Let us now focus on the effect of a magnetic field. The metamagnetic transition in \csrea\ occurs approximately at 6~T for a magnetic
field along the $c$-direction and at 3~T for B$\parallel$\textit{a,b} \cite{nakatsuji03,balicas}. The even stronger in-plane anisotropy
between the two orthorhombic axes (diagonals in our notation) is not relevant in our experimental geometry. In Figure \ref{fig3} we
present the data for B$\parallel$(010). There is a drastic change at the metamagnetic transition field ($B_\text{MM}\approx 3\,\text{T}$),
which corresponds to an only low energy scale $g\mu_BB\approx0.17\,\text{meV}$: below the transition, we observe the incommensurate AFM
fluctuations. Above, the response is dominated by a broad paramagnon signal around the 2D FM zone center (0,0,1.6). The data can be
described in a very simple model of two contributions: an AFM one (taken to have always the same shape as at B=0) and a FM one. It is
clearly seen that the fundamental change takes place between 2 and 4T, i.e.\ at the metamagnetic transition. In the scans, a small AFM
contribution seems to persist to higher fields and to be suppressed only far above the transition. The energy scans at the FM zone center
(Fig.\ \ref{fig3}b) confirm that upon applying a magnetic field the FM response is enhanced and that its spectral weight shifts to higher
energies at fields above the transition. The transition is further seen in the field dependence of the intensity at fixed energy (Fig.\
\ref{fig3}c,d). Low-energy paramagnon fluctuations seem to govern the thermodynamics of the metamagnetic transition. Their enhancement at
the critical field (Fig.~\ref{fig3}d) may explain the observation that the electronic specific heat passes a maximum at the transition
\cite{baier}, because especially the low-energy magnetic fluctuations give a large contribution to the electronic specific heat.

The quantitative analysis of the paramagnon signal in the B$\parallel$(010) configuration is difficult. Due to the low vertical resolution
of the focusing spectrometer, the dispersion along \textit{b*} is averaged. Furthermore, the measured susceptibility consists of a
superposition of several components. At \textbf{Q}=(0,0,L) we measure the sum of the in-plane-components $\chi''_{ab}$. Because the field
is along \textit{b*}, $\chi''_{ab}$ splits into two components, $\chi^{\parallel}$ and $\chi^{\perp}$, parallel and transverse to the
field. The transverse susceptibility represents electronic spin-flip processes and, therefore, senses a Zeeman-type energy gap, $g\mu_BB
\approx 1$~meV for a field of 10~T, in addition to the spin-orbit coupling induced anisotropies, whereas $\chi^{\parallel}$ only senses the
anisotropy gap.

The quantitative analysis becomes simpler when the field is applied along the $c$ axis, see Fig.\ \ref{fig5}. The low vertical resolution
now averages along the $c$ direction, where there should not be any significant magnetic dispersion. Again there is a drastic change at
the metamagnetic transition ($B_\text{MM}=6T$), see Fig.\ \ref{fig5}a,b and the zero field data taken under the same conditions in Fig.\
\ref{fig2}. At 10~T, there is clear evidence for a well defined magnon mode. This mode disperses outwards from (1,0,0), rapidly broadens
and dies away with increasing frequency. It can be described by a quadratic dispersion (Fig.\ \ref{fig5}c) similar to a conventional
ferromagnet: $\hbar \omega =g \mu _B B_{\text{eff}} + D q^2$ with a stiffness constant of 60\ meV\AA$^2$. Since the external field
superposes the anisotropy terms, an effective field enters the dispersion relation. Assuming that the single-ion anisotropies nearly
average out in this geometry, we set $B_\text{eff}\approx B_\text{external}$. The dispersion of the magnon unambiguously proves that the
high-magnetization phase does not only arise from the magnetic polarization of the spins, but has to be considered as a true FM state with
an intrinsic FM interaction stabilized by the magnetic field. In contrast, the FM correlation seems to be efficiently suppressed in the
low-temperature low-field phase due to the orbital effects \cite{kriener,balicas}.

We mention that an enhancement of magnetic scattering near FM q-vectors has also been observed in the metamagnetic transition in the heavy
fermion compound CeRu$_2$Si$_2$ \cite{cerusi}, but a well-defined magnon mode has not been established in any metamagnetic transition so
far.

In conclusion, our results indicate that a magnetic field corresponding to low electronic energy scales causes a fundamental change of the
magnetic correlations in \csrea . Below the metamagnetic transition, the interactions are AFM with scattering contributions from several
incommensurate wave vectors.  In particular, there is a weak signal near (0.3,0.3,q$_L$), i.e.\ very close to the position of the dominant
nesting signal in \sro . Magnetic correlations in \csrea\ fundamentally change upon increase of either temperature or magnetic field. In
both cases we find strong paramagnon scattering, unambiguously proving the different character of these states. The metamagnetic
transition seems to arise from the competition of incommensurate and ferromagnetic instabilities. In the high-magnetization phase we even
find a well-defined magnon mode indicating the dominance of the FM interaction and the intrinsic FM character of the high-field state.

\paragraph{Acknowledgements.} Work at Universit\"at zu K\"oln was supported by the
Deutsche Forschungsgemeinschaft through the
Sonderforschungsbereich 608.


\begin{thebibliography}{}
\bibitem{grigera} S.A. Grigera \textit{et al.}, Science \textbf{294}, 329 (2001); \; R.S. Perry \textit{et al.},
Phys. Rev. Lett. \textbf{86}, 2661 (2001); \; S.A. Grigera \textit{et al.}, Science \textbf{306},
1154 (2004).
\bibitem{nakatsuji00} S. Nakatsuji and Y. Maeno, Phys. Rev. Lett. \textbf{84}, 2666 (2000); \; S. Nakatsuji and Y. Maeno, Phys. Rev.
B \textbf{62}, 6458 (2000).
\bibitem{nakatsuji03} S. Nakatsuji \textit{et al.}, Phys. Rev. Lett. \textbf{90}, 137202 (2003).
\bibitem{kriener} M. Kriener \textit{et al.}, Phys. Rev. Lett. \textbf{95}, 267403 (2005).
\bibitem{balicas} L. Balicas \textit{et al.}, Phys. Rev. Lett. \textbf{95}, 196407 (2005).
\bibitem{joerg} J. Baier \textit{et al.}, cond-mat 0610769 (2006).
\bibitem{lovesey} see for instance S.W. Lovesey, \textit{Theory of neutron scattering from condensed matter, Vol. 2} (Clarendon, Oxford,
1984), or T. Moriya, \textit{Spin fluctuations in itinerant electron magnetism} (Springer, 1985).
\bibitem{friedtprl} O. Friedt \textit{et al.}, Phys. Rev. Lett. \textbf{93}, 147404 (2004).
\bibitem{capogna} L. Capogna \textit{et al.}, Phys. Rev. B \textbf{67}, 012504 (2003).
\bibitem{stone} M.B. Stone \textit{et al.}, Phys. Rev. B \textbf{73}, 174426 (2006).
\bibitem{friedtprb} O. Friedt \textit{et al.}, Phys. Rev. B \textbf{63}, 174432 (2001).
\bibitem{sidis} Y. Sidis \textit{et al.}, Phys. Rev. Lett. \textbf{83}, 3320 (1999); \\ M. Braden \textit{et al.}, Phys. Rev. \textbf{B} {66}, 064522
(2002).
\bibitem{pol-srruo}\ M. Braden \textit{et al.}, Phys. Rev. Lett. \textbf{92},
097402 (2004).
\bibitem{band} C. Bergemann \textit{et al.}, Adv. in Phys. \textbf{52}, 637 (2003); \\ I. Mazin and D. Singh, Phys. Rev. Lett. \textbf{79}, 733
(1997).
\bibitem{anisimov} V.I. Anisimov \textit{et al.}, Eur. Phys. J. B \textbf{25}, 191 (2002).
\bibitem{gukasov} A. Gukasov \textit{et al.}, Phys. Rev. Lett. \textbf{89}, 087202 (2002).
\bibitem{wang} S.C. Wang \textit{et al.}, Phys. Rev. Lett. \textbf{93}, 177007 (2004).
\bibitem{baier} J. Baier \textit{et al.}, Physica B \textbf{378}, 497 (2006).
\bibitem{cerusi} M. Sato \textit{et al.}, J. Phys. Soc. Jpn. \textbf{12}, 3418 (2004).

\end{thebibliography}
\end{document}